\documentclass[fleqn,usenatbib]{mnras}

\usepackage{newtxtext,newtxmath}
\usepackage{color}

\usepackage[T1]{fontenc}

\DeclareRobustCommand{\VAN}[3]{#2}
\let\VANthebibliography\thebibliography
\def\thebibliography{\DeclareRobustCommand{\VAN}[3]{##3}\VANthebibliography}

\usepackage{graphicx}	
\usepackage{amsmath}	
\usepackage{makecell}

\usepackage{hyperref}
\usepackage{booktabs}
\usepackage{amsmath}
\usepackage{xcolor}

\title[eBOSS ELG abundance matching]{Abundance matching analysis of the emission line galaxy sample in the extended Baryon Oscillation Spectroscopic Survey}

\author[Lin et al.]{\parbox{\textwidth}{
Sicheng Lin,$^{1}$\thanks{E-mail: sicheng@nyu.edu}
Jeremy L. Tinker,$^{1}$ 
Michael R. Blanton,$^{1}$
Hong Guo, $^{2}$
Anand Raichoor,$^{3}$
Johan Comparat,$^{4}$
Joel R. Brownstein$^{5}$
}\\
$^{1}$Center for Cosmology and Particle Physics, Department of Physics, New York University, 726 Broadway, New York, NY 10003, USA\\
$^{2}$Key Laboratory for Research in Galaxies and Cosmology, Shanghai Astronomical Observatory, Shanghai 200030, China\\
$^{3}$Institute of Physics, Laboratory of Astrophysics, Ecole Polytechnique Fédérale de Lausanne (EPFL), Observatoire de Sauverny, 1290 Versoix, Switzerland\\
$^{4}$Max-Planck-Institut f\"{u}r extraterrestrische Physik (MPE), Giessenbachstrasse 1, D-85748 Garching bei M\"unchen, Germany\\
$^{5}$Department of Physics and Astronomy, University of Utah, 115 S. 1400 E., Salt Lake City, UT 84112, USA
}

\date{Accepted 2022 September 23. Received 2022 September 23; in original form 2022 January 21}
\pubyear{2022}

\begin{document}
\label{firstpage}
\pagerange{\pageref{firstpage}--\pageref{lastpage}}
\maketitle

\begin{abstract}
We present the measurements of the small-scale clustering for the emission line galaxy (ELG) sample from the extended Baryon Oscillation Spectroscopic Survey (eBOSS) in the Sloan Digital Sky Survey IV (SDSS-IV). We use conditional abundance matching method to interpret the clustering measurements from $0.34h^{-1}\textrm{Mpc}$ to $70h^{-1}\textrm{Mpc}$. In order to account for the correlation between properties of emission line galaxies and their environment, we add a secondary connection between star formation rate of ELGs and halo accretion rate. Three parameters are introduced to model the ELG [OII] luminosity and to mimic the target selection of eBOSS ELGs. The parameters in our models are optimized using Markov Chain Monte Carlo (MCMC) method. We find that by conditionally matching star formation rate of galaxies and the halo accretion rate, we are able to reproduce the eBOSS ELG small scale clustering within 1$\sigma$ error level. Our best fit model shows that the eBOSS ELG sample only consists of $\sim 12\%$ of all star-forming galaxies, and the satellite fraction of eBOSS ELG sample is 19.3\%. We show that the effect of assembly bias is $\sim20\%$ on the two-point correlation function and $\sim5\%$ on the void probability function at scale of $r\sim 20 h^{-1}\rm Mpc$.
\end{abstract}



\section{Introduction}
\label{introduction}
In modern cosmology, galaxies form and evolve within dark matter halos. In the early universe, the initial fluctuations of the matter field grow and evolve under gravity. Baryonic matter falls into dark matter halos and forms stars and galaxies. During the evolution of the universe, dark matter halos form through accretion and mergers \citep[e.g., ][]{lacey1993}. At the same time, galaxies within dark matter halos form stars and grow through mergers as well. The details of the growth history and merger history depend on cosmological parameters and the physics of galaxy formation. It is important to study the relationship between galaxies and dark matter halos, because it helps us understand the properties of dark matter and constrain cosmological parameters. This paper constructs an empirical model of the newly completed eBOSS emission line galaxy (ELG) sample and its host dark matter halos.

There are many ways to model the relationship between galaxies and dark matter halos (for a recent review, see \citealt{Wechsler2018}). The most physical methods take into consideration the complicated physics of galaxy formation by using hydrodynamic simulations to simulate the galaxy formation process (an example of recent simulations see the EAGLE project; \citealt[][]{Schaye2015}). Semi-analytical methods \citep[SAMs; ][]{White1991, Kauffmann1993, Somerville1999, Cole2000} aim to estimate the evolution of gas and star formation without explicitly incorporating hydrodynamics. Although closer to physical realism, these models are generally very computationally expensive. Fully empirical methods use statistical ways to connect properties of galaxies with properties of dark matter halos. Subhalo abundance matching \citep[SHAM;][]{Kravtsov2004, Vale2004, Conroy2006}, in its most basic form, assumes that each (sub)halo hosts a galaxy and there is a monotonic relation between galaxy mass and (sub)halo mass. A successful SHAM model requires high-resolution dark matter simulations to locate the substructure inside host halos because the subhalos are naturally the local environment for hosting satellite galaxies. Another statistical model, the halo occupation distribution \citep[HOD;][]{Peacock2000,Seljak2000,Benson2000,zheng2004,scoccimarro2001}, models the probability $P(N|M)$ of finding $N$ galaxies of a certain type in a dark matter halo with halo mass $M$. The spatial and velocity distribution of galaxies are determined by the halo mass profiles. HOD modeling also utilizes high-resolution N-body simulations to accurately identify halos and interpret galaxy clustering measurements, but it doesn't require the identification of substructure inside halos. 

In this paper we use abundance matching to model eBOSS ELGs. Compared with physical methods, SHAM circumvents the complex physics of galaxy formation and utilizes dark matter halo properties such as halo mass function, halo bias, halo mass profiles, etc, to interpret galaxy clustering statistics. This method is relatively computationally inexpensive and easy to implement once high-resolution simulations are given. Because of the above reason, SHAM has become a popular tool to study galaxy-halo relation in recent years, along with HOD. They all provide statistical link between dark matter halos and galaxies, and have been widely used in recent spectroscopic surveys \citep[e.g. ][]{Avila2020, Yu2022}. Some efforts are made to add more flavor into classical HOD for more physical understanding of galaxy formation. For example, \citet{Hearin2016} introduced a method to model assembly bias in HOD. Compared with HOD, SHAM naturally provides an opportunity to account for (sub)halo evolution history and model galaxies' dependency on their environment. This can help us gain insights on physics of galaxy formation. 

Despite its simple assumptions about the galaxy-halo connection, abundance matching has proven successful in simultaneously modeling the stellar mass function (SMF) as well as the two-point correlation function \citep[see e.g.][for results on SDSS, DEEP2 and EAGLE simulations]{Conroy2006,Chaves2016,Rodr2016}. The first abundance matching implementations related the stellar mass monotonically with the halo mass. However, because the masses of subhalos are sensitive to tidal stripping, it is more plausible to relate the stellar mass with properties of subhalos at the time they merged into their host halos. Several proxies are studied and compared with each other later, and it was shown that using the peak maximum circular velocity yields better result than using halo mass \citep{Reddick2013}. Another puzzle piece in the abundance matching model is the scatter in the stellar mass at a given halo property. It describes the variance in stellar mass that cannot be explained by the halo property that we choose. Models with different scatter can produce the same SMF, but with different two-point galaxy clustering statistics as well as the stellar-to-halo mass relations (SHMR). Normally the scatter is assumed to be log-normal and independent of halo mass \citep{Yang2009, behroozi2010}, although some recent studies have shown that the scatter may be higher at lower halo masses \citep{Cao2020, Taylor2020, Tinker2020}. 

While simple abundance matching models have proven successful, the actual galaxy-halo connection is more complex than the relationship between a single galaxy property with a single halo property. One phenomenon from observations is the galaxy bimodality, which separates the galaxies into a red sequence (usually old galaxies) and a blue sequence (usually young galaxies) in color space. The red galaxies tend to reside in overdense regions while the blue galaxies appear more often in underdense regions. This color-dependent distribution of galaxies can be revealed in the two-point statistics. \citet{Hearin2013} proposed a method called ``conditional abundance matching'' for incorporating secondary galaxy properties into such models. At fixed stellar mass (or any other primary halo property) a secondary galaxy property is abundance matched to a secondary halo property. Secondary halo properties, such as halo age and concentration, strongly influence the clustering of halos at fixed mass \citep[see, e.g.][and references therein]{Wechsler2018}. This is called assembly bias. Using this method, \citet{tinker2017b, tinker2018b} and \citet{Zu2016, Zu2018} found no strong evidence for secondary correlations between halo properties and galaxy color. This method has also been used to explore various second galaxy properties such as star formation rate \citep{watson2015}, galaxy morphology \citep{tinker2018} and galaxy size \citep{Hearin2017}. Conditional abundance matching is an extremely useful tool to study the relationship between galaxies and dark matter halos. In this paper we adopt a similar approach as \citet{tinker2018} to model star-forming galaxies.

As galaxy redshift surveys push to higher redshifts, ELGs have become the preferred tracer of the dark matter density field. The strong emission lines and the abundance of ELGs at redshift $z\sim 1$ make them suitable to trace dark matter and constrain cosmological models. The SDSS-IV/eBOSS survey \citep{dawson2016} is a newly completed spectroscopic survey that measures redshift of $\sim$174,000 ELGs at redshift range $0.6<z<1.1$. It is so far one of the largest ELG sample to measure large-scale clustering. Future surveys that use ELGs as cosmological tracers include DESI \citep{DESI2016a, DESI2016b}, PFS \citep{takada2014}, 4MOST \citep{dejong2014}, Euclid \citep{Laureijs2011}, NGRST \citep[formerly known as WFIRST][]{Spergel2013}, etc. Understanding the ELG-halo connection is beneficial to not only future surveys that heavily use ELGs, but also  studies of quenching mechanisms of star formation, because the cosmic star formation peaks at around $z\sim 2$ \citep{Madau2014}. The framework we developed here should be able to be applied to other surveys to inform target selection and large scale structure analysis.

The purpose of this paper is to build a physically-motivated model for the star forming galaxies with the possibility of galaxy assembly bias taken into account. To do so, we try to create physical and realistic ELG mocks with minimum parameters involved. Our model assumes that there are two connections between star forming galaxies and dark matter halos. The first is the monotonic relationship between the peak maximum circular velocity of halos and the stellar mass. The second one is the monotonic relationship between the halo accretion rates and the star formation rates of galaxies. The second relationship which we modulate, is intended to link the physical process of star formation within galaxies with halo assembly history \citep{watson2015, tinker2018}. This provides physical insights about the galaxy formation to explain the ELG clustering, satellite fractions, etc.

This paper is structured as follows. In \S \ref{sec: data} we describe the eBOSS/ELG data sample. The methods of measuring small scale clustering and the observed [$\rm OII$]3727 luminosity function are described in \S \ref{subsec: LOII} and \S \ref{subsec: ssc}, respectively. \S \ref{sec: method} describes the Multidark Planck simulation as well as the details of our abundance matching model. We present our result in \S \ref{sec: result} and interpret the clustering of eBOSS ELG sample. In \S \ref{sec: conclusion} we describe our key findings. Throughout the paper we use a spatially flat $\Lambda$CDM cosmology with $\Omega_M = 0.307, \Omega_L = 0.693, h_0 = 0.678, \Omega_b = 0.0482, n_s = 0.96$ and $\sigma_8=0.823$. It's also the same cosmology that is used to run the MultiDark Planck 2 simulation.

\section{Data}\label{sec: data}
In this section we describe the eBOSS ELG data sample that we use in this paper. Our model will reproduce the distribution of $[\rm OII]$ luminosity in the eBOSS ELG sample, as well as the small-scale projected two-point clustering of ELGs.


\subsection{eBOSS ELG sample}\label{subsec: ebossELG}
The extended Baryon Oscillation Spectroscopic Survey (eBOSS) is a newly completed six-year program \citep[2014 -- 2020; ][]{dawson2016} within the Sloan Digital Sky Survey IV \citep[SDSS-IV; ][]{blanton2017}. It uses the Sloan Foundation 2.5-m Telescope at the Apache Point Observatory \citep{Gunn2006} with the same BOSS fiber spectrographs \citep{Smee2013} to conduct observations. The goal of the eBOSS is to constrain cosmological models by measuring the large-scale structure of the universe, and it has successfully measured the cosmological distance scale at  the percent level at four different redshifts \citep{eboss2020}. The eBOSS survey uses three different tracers to maximize redshift coverage: emission line galaxies (ELGs), luminous red galaxies (LRGs) and quasars. The eBOSS ELG program utilizes the DECam Legacy Survey \citep[DECaLS; ][]{Dey2018} to perform target selection, because it has deeper $grz-$band photometry than the SDSS imaging \citep{raichoor2017}. We use the final eBOSS ELG catalog from the SDSS-IV Data Release 16 (DR16) in this paper. It contains 173,736 reliable spectroscopic measurements of redshift between $0.6<z<1.1$, covering a total effective area of 727 deg$^2$ \citep{raichoor2021}.

\subsection{Observed [$\rm OII$] luminosity function}
\label{subsec: LOII}

All abundance matching models should begin with observational measurements of the luminosity function or the mass function of the sample being modeled. Although we are modeling a sub-class of galaxies, this is still a critical piece of constructing our model. From the eBOSS ELG sample we measure the [$\rm OII$]3727 luminosity function, which quantifies the abundance of the ELG at a given [$\rm OII$] luminosity. The reason we choose [$\rm OII$] luminosity function instead of stellar mass function is that [$\rm OII$] flux can be obtained directly from the spectra, while stellar masses of ELGs are measured in indirect ways such as spectral energy distribution (SED) fitting.

The [$\rm OII$] flux of each ELG is extracted using the \texttt{IDLSPEC2D} eBOSS pipeline. The [$\rm OII$] luminosity of each ELG is then calculated as 

\begin{equation}
    L[\textnormal{OII}] = f[\textnormal{OII}] \times 4\pi D_L^2(z),
\end{equation}
where $f[\rm{OII}]$ is the [$\rm OII$] flux and $D_L(z)$ is the luminosity distance at redshift $z$. We show in Fig. \ref{fig: LOII} the observed [$\rm OII$] luminosity of the eBOSS ELG sample, measured in the luminosity range of $10^{40.5} \rm{erg~s^{-1}}$ $< L[\rm OII] < 10^{42.5} \rm erg~s^{-1}$ with logarithmic bins of $\Delta \log L[\rm OII]=0.05$ dex. The errors are estimated using a bootstrap method. The Gaussian shape of the measured luminosity function is a product of the standard schechter form for such data and the eBOSS ELG selection function, which becomes inefficient at low luminosities.

\subsection{Small-scale clustering}\label{subsec: ssc}
Besides the one-point statistics of galaxies, i.e., the galaxy luminosity function, another important tool to study the statistics of galaxies is the two-point correlation function (2PCF) $\xi(r)$. It describes the excess probability to find a galaxy pair with separation $r$ over a random galaxy field \citep{Peebles1980}. Quantifying the 2PCF of galaxies is useful because it distinguishes models with different galaxy bias at scales of $r\gtrsim 1\rm Mpc$ and constrains the fraction of satellite galaxies at scales of $r \lesssim 1 \rm Mpc$. Bias $b$ is defined as the ratio of clustering relative to clustering of dark matter, $b^2=\xi/\xi_{\rm DM}$. The bias of dark matter halo is not sensitive to the halo mass for small halos, however it is very sensitive to the halo mass for halos with higher mass. The large scale bias provides information on the overall halo mass scale for the galaxy sample, while the small scale clustering constrains the fraction of the sample that are satellite galaxies.

We use the Landy \& Szalay estimator \citep{Landy1993} to measure the projected 2PCF $w_p(r_p)$ for the eBOSS ELG sample. We correct the systematic effects on the clustering using the weighting scheme described in \citet{raichoor2021}. Each individual ELG is weighted by
\begin{equation}
    w_{\textnormal{ELG}} = w_{\textnormal{FKP}}w_{\textrm{sys}}w_{\textrm{cp}}w_{\textrm{noz}},
\end{equation}
where $w_{\textnormal{FKP}},w_{\textrm{sys}},w_{\textrm{cp}}$ and $w_{\textrm{noz}}$ are the FKP weights \citep{fkp1994}, the imaging systematic weights, the close-pair weights, and the redshift failure weights, respectively. We refer readers to \citet{raichoor2021} for details of ELG systematic weights. We choose 17 logarithmic bins from $0.34h^{-1}\textrm{Mpc}$ to $70h^{-1}\textrm{Mpc}$, covering both the one-halo term and two-halo term ranges. The errors are estimated with the jackknife resampling technique using 25 subsamples. Our measurements are presented in Fig. \ref{fig: wp}.

Estimating the covariance matrix directly from the data is noisy given the limited size of ELG sample. Thus to quantify the uncertainties in ELG clustering, we utilize the GLAM-QPM mocks from \citet{Lin2020}. This set of mocks model the linear bias of eBOSS ELG sample at the scales we are interested. We first compute the covariance matrix $C_{ij} = \textrm{cov}(\xi_i, \xi_j)$ from 2000 GLAM-QPM mocks, where $\xi_i$ and $\xi_j$ are the clustering measurements at the $i^{\textrm{th}}$ and $j^{\textrm{th}}$ bin, respectively. Then we up-scale it by the jackknife errorbar of $\xi_{\rm ELG}$,
\begin{equation}
    C'_{ij} = \frac{C_{ij}}{\sqrt{C_{ii}C_{jj}}} e_i e_j,
\end{equation}
where $e_i$ and $e_j$ are the errors of the clustering measurements at the $i^{\textrm{th}}$ and $j^{\textrm{th}}$ bin, respectively.

\section{Theoretical Model}\label{sec: method}
Abundance matching is a technique to describe the relationship between galaxies and dark matter halos. As described in \S \ref{introduction}, a scheme that only maps one property of galaxies with one property of halos may not be sufficient, especially for star forming galaxies like ELGs. So we adopted a conditional abundance matching model which not only matches primary properties, but also conditionally matches the star formation rate with the halo accretion rate. Although this model is physically motivated and supported by local observations, we test the impact of removing this correlation as well. We will describe the details of our method in the following sub-sections. Here we briefly summarize our abundance matching procedure of modeling eBOSS ELGs in the following steps:

\begin{itemize}
    \item In order to generate a mock galaxy catalog for all galaxies, we apply the abundance matching technique with a 0.15 dex scatter in stellar mass at fixed $V_{\rm peak}$ of halo.
    \item To separate star-forming galaxies from passive galaxies, we utilize the quenched fraction from \citet{Tinker2013} for both central and satellite galaxies. We randomly select actively star-forming galaxies from the overall mock galaxy catalog to match the observed fractions.
    \item A star forming main sequence fitting model is used to estimate star formation rate (SFR) of each mock galaxy, with a 0.3 dex scatter. We then perform a rank matching of the SFR with the accretion rate of the host halo at fixed stellar mass.
    \item The $[\rm OII]$ luminosity of each galaxy in the mock catalog is estimated using each galaxy's SFR with an empirical correction. We further use a parameterized selection function based on the  $[\rm OII]$ luminosity in order to mimic the target selection of eBOSS ELG. This step introduces three free parameters. The description of the parameters is listed in Table \ref{table: parameters}.
    \item For each parameter set, we construct the $[\rm OII]$ luminosity function and the two-point correlation function from the mock catalogs. The likelihood of the luminosity function as well as the two-point correlation function are used in the MCMC chain to optimize the parameters. 
\end{itemize}

\begin{table}
\begin{center}
\caption{Parameters in our abundance matching model.}
\begin{tabular}{@{}ll@{}}
\toprule
parameter  & value \\ \midrule
$\sigma(\log M_\star|\log V_{\rm peak})$      & 0.15 dex \citep{behroozi2010} \\
$\sigma(\log \textrm{SFR}|\log M_\star)$     & 0.3 dex \citep{Noeske2007}    \\
$\sigma(\log L[\textrm{OII}]|\log \textrm{SFR})$ &  0.12 dex (best-fit)  \\
$\log L_{crit}$ &  41.7 dex (best-fit)  \\ 
$\sigma_{\log L}$ &  0.63 dex (best-fit)\\ \bottomrule
\end{tabular}
\label{table: parameters}
\end{center}
\end{table}

\subsection{Dark matter simulations}\label{subsec: DMsimulation}

In this paper we use the dark matter halo catalogs from the MultiDark Planck 2 simulation \citep[MDPL2\footnote{https://www.cosmosim.org/cms/simulations/mdpl2/}; ][]{prada2012, klypin2016}. The MDPL2 simulation adopts a flat $\Lambda$CDM cosmology with cosmological parameters described in \S \ref{introduction}, and is run by the \texttt{L-GADGET-2} code, which is a variant of cosmological code \texttt{GADGET-2} \citep{Springel2005}. The box size of the simulation is $1h^{-1}\textrm{Gpc}$, and the mass resolution is $1.51\times10^9 h^{-1}M_\odot$. We choose the simulation output at redshift $z=0.819$, roughly equal to the ELG effective redshift $z_{\textrm{eff}}=0.845$. Halos are found using the \texttt{RockStar} algorithm \citep{Behroozi2013}. The peak maximum circular velocity $V_{\rm peak}$ and the peak halo mass $M_{\rm peak}$ are determined throughout the merger history. The attributes in the dark matter halo catalogs that are used in this paper and their definitions are presented in Table \ref{table: attributes}. The reason we choose MDPL2 simulation is two folds. First it has high mass resolution, allowing us to locate halos at low mass end (around $10^{11} h^{-1}M_\odot$). This is crucial in our study because ELGs are young galaxies that occupy lower-mass halos. Second, the volume of the simulation is large enough compared with eBOSS ELG footprint for us to get a good comparison to the data.

\begin{table}
\begin{center}
\caption{Attributes in MDPL2 simulation that are used in this paper.}
\begin{tabular}{@{}lll@{}}
\toprule
attribute name   & unit & definition \\ \midrule
$M_{\textrm{peak}}$      & $h^{-1}M_\odot$ & the peak halo mass over accretion history \\
$V_{\textrm{peak}}$     & $\rm km/s$ & \thead[l]{the peak maximum circular velocity\\over accretion history}    \\
$M_{200b}$ & $h^{-1}M_\odot$ & halo mass enclosed within overdensity $200\rho_b$   \\
$\textrm{accRate}_{M_{\textrm{peak}}}$ & $h^{-1}M_\odot/yr$ & growth rate of $M_{\textrm{peak}}$    \\ \bottomrule
\end{tabular}
\label{table: attributes}
\end{center}
\end{table}

\subsection{Abundance Matching Model}
\label{subsec:SHAM}

\subsubsection{Stellar Mass Function of All Galaxies}
Since our abundance matching approach uses the stellar mass $M_\star$ as the galaxy property to match with $V_{\rm peak}$ of halos, it requires the abundance of galaxies as a function of their stellar masses. We use the fitted SMF from \citet{ilbert2013} at redshift $0.8<z<1.1$.
Note that the SMF from \citet{ilbert2013} is for all galaxies. The ELG sample from eBOSS that has been selected using color cuts, however, is a subset of the full sample. We will tackle this problem later in \S \ref{subsec:QF}.


\subsubsection{SubHalo Abundance Matching}
\label{subsubsec:SHAM}
We follow the procedure of \citet{behroozi2010} in order to match $V_{\rm peak}$ and $M_\star$ according to their abundance. First we deconvolve the scatter from the true SMF to get a `direct' SMF, which is used to match with $V_{\rm peak}$ directly. We assign a constant 0.15 dex scatter in $M_\star$ at a given $V_{\rm peak}$ \citep{behroozi2010}. Second, we perform the abundance matching using the direct SMF with $V_{\rm peak}$ from the MDPL2 halo catalog, and produce a mock galaxy catalog. At last we re-add the log-normal scatter to the stellar mass in the mock galaxy catalog, yielding a match to the input SMF from \citet{ilbert2013}.

\subsubsection{Quenched Fraction}
\label{subsec:QF}
The [OII] doublet emission lines in the spectrum of ELGs indicate that they are strongly star-forming. However, the mock galaxy catalog that we built using SMF from \citet{ilbert2013} contains all types of galaxies. In order to distinguish the young, blue and star-forming galaxies with the old, red and quiescent galaxies, we apply the quenched fraction at $z=0.88$ from \citet{Tinker2013}. The quenched fraction $f_q$ is defined as the fraction of galaxies that has no star formation (i.e., is quenched). In \citet{Tinker2013}, the separation of quenched and star-forming galaxies is based on the standard UVJ diagram. For the galaxy mass range we consider, the quenched fraction monotonically increases with stellar mass. The physical mechanisms of star formation quenching can be different for central galaxies and satellite galaxies. For central galaxies, the quenching is likely due to halo mergers or AGN feedback, whereas for satellite galaxies it is more likely due to tidal stripping from the host halos and ram pressure stripping from intracluster medium. Hence the relationship between the quenched fraction and the stellar mass is different for central galaxies and satellite galaxies. Being able to distinguish the different quenched fractions for centrals and satellites is important, because it could have impacts on the small-scale galaxy clustering. Getting the right amount of satellites helps us constrain the clustering at scales of $r\lesssim 1h^{-1}\rm Mpc$. 

We apply the quenched fraction for both centrals and satellites from \citet{Tinker2013} in the stellar mass range $10^{9.7}h^{-1}M_\odot \leq M_\star \leq 10^{11.2}h^{-1} M_\odot$, and we perform a linear interpolation between each pair of adjacent data points ($\log f_q$ vs. $\log M_\star$) as well as a linear extrapolation for the low mass end  $M_\star<10^{9.7}h^{-1}M_\odot$ and the high mass end $M_\star>10^{11.2}h^{-1} M_\odot$. If the linear extrapolation of quenched fraction exceeds 1, we truncate the value to 1. We then take the mock galaxy catalog from \S \ref{subsubsec:SHAM} and randomly remove galaxies according to the quenched fraction based on their stellar masses and whether they are central galaxies. In this step we take a random subset of halos to host quenched central galaxies. This matches $z=0$ observations that the quenched fraction of centrals is independent of environment \citep{tinker2017b, tinker2018b}, it only depends on the stellar mass in our case. 

One alternative way to compute the quenched fraction is performing a linear fit of $\log f_q$ vs. $\log M_\star$ in \citet{Tinker2013} instead of linear interpolation/extrapolation as what we just described. We performed a sensitivity test and did not see a significant difference in the galaxy clustering between these two methods.

\subsubsection{Conditional Abundance Matching}
\label{subsubsec: CAM}
So far we have constructed a mock galaxy catalog for star forming galaxies using a single connection between stellar mass and peak maximum circular velocity of halos. However, the properties of star forming galaxies are correlated with their environment, and the assembly bias and galaxy formation has not yet been incorporated in the current model. Here we link the SFR of galaxies with the accretion rate of halos at a fixed stellar mass, using the conditional abundance matching \citep[CAM;][]{Hearin2013, Hearin2014} approach employed in \citet{tinker2018}. In this way, for a given stellar mass, the galaxy which has the highest SFR resides in the halo which has the highest growth rate \citep{tinker2018}. This is the maximal impact of assembly bias. We also show results when this correlation is removed completely, thus showing the range of possible models. In order to calculate SFR for each galaxy in the mock catalog, we use the star forming main sequence fitting model from \citet{Lee2015}:
\begin{equation}
S = S_0 - \log \left[1+\left(\frac{M_\star}{M_0}\right)^{-\gamma}\right],
\end{equation}
where $S=\log(\textnormal{SFR})$, $S_0=1.35 \log(M_\odot \textrm{yr}^{-1})$ , $M_0=10^{9.96} M_\odot$, and $\gamma=1.28$. We use the best-fit parameters for redshift range $0.78<z<0.93$. We add a 0.3 dex scatter to $S=\log(\textnormal{SFR})$ \citep{Noeske2007}, and match SFR with halo mass accretion rate given a stellar mass bin.

\subsubsection{Estimating $L[\rm OII]$ }
In \S \ref{subsubsec: CAM} we use CAM to link the SFR of galaxies with the growth rate of host halos, under the assumption that galaxies and halos co-evolve. Generally speaking, a galaxy with a higher SFR will have a stronger emission line in its spectrum, and it is more likely for us to observe. But the eBOSS ELG target selection was conducted using a color cut in the observed frame $g-r$ vs. $r-z$ space, as well as a cut on the $g$-band magnitude. The color and the SFR of a galaxy are not perfectly correlated with each other. For a galaxy with active star formation, it is gas rich and can appear red due to dust attenuation. In addition, the redshifts from eBOSS ELGs were measured using $[\rm OII]$ doublet, and the $[\rm OII]$ luminosity is strongly affected by dust. Moreover, the $[\rm OII]$ doublet emission lines are produced from metal ions by collisional excitation, and is thus correlated with metallicity. Given the reasons above, the SFR of a galaxy is not a direct proxy for us to select eBOSS ELG targets from the mock catalog. Instead, we estimate the $[\rm OII]$ luminosity from the SFR in the mock catalog using an empirical correction from \citet{Gilbank2010},

\begin{equation}
\label{eq: LOII}
L[\textrm{OII}] = (2.53\times 10^{40}\textrm{erg~ s}^{-1})\cdot \textrm{SFR}\cdot(a\tanh[(\log M_\star - b)/c] + d),
\end{equation}
where $a = -1.424, b=9.827, c=0.572, d=1.700$. And we introduce a scatter in $\log L[\rm OII]$ given a SFR, $\sigma(\log L[\rm OII] | \log SFR)$, as a free parameter in our model.

To mimic the target selection of eBOSS ELGs, an erf function is used to cut mocks based on $L[\rm OII]$. The magnitude and color limits in the ELG target selection will not produce a hard threshold in $L[\rm OII]$, and the erf function with a transition width is intended to mimic this effect. Eq. \ref{eq: erf} describes the probability of a galaxy in the mock with $[\rm OII]$ luminosity $L$ being selected as an eBOSS ELG target. This introduces two free parameters: $L_{crit}$ is the luminosity that has 50\% probability for an ELG to be selected, and $\sigma_{\log L}$ is the transition width of the selection function. The coefficient $\eta$ is added to describe the incompleteness of the eBOSS ELG sample. Note that the coefficient $\eta$ to the erf function is not a free parameter as it is constrained by the eBOSS ELG number density.

\begin{equation}
\label{eq: erf}
    P(\log L) = \eta \left[1 + \textrm{erf}\left(\frac{\log L - \log L_{crit}}{\sigma_{\log L}}\right)\right].
\end{equation}

\subsection{The model parameters and MCMC}
\label{subsec: MCMC}
As described in \S \ref{subsec:SHAM}, the 3 free parameters in our abundance matching model are $\sigma(\log L[\rm OII] | \log SFR)$, $L_{crit}$ and $\sigma_{\log L}$. Our implementation of the model is able to quickly produce mock galaxy catalogs and accurately compute small scale clustering for a given parameter set, making it ideal to be constrained using MCMC method. For each trial parameter set in the MCMC chain, we calculate a likelihood of the observed $L[\rm OII]$ luminosity function and $w_p(r_p)$. We use the \texttt{emcee} package \citep{emcee2013}, which implements the affine-invariant ensemble sampler \citep{Goodman2010}, to perform the MCMC analysis.

For the $L[\rm OII]$ luminosity function, we use KL-divergence (also called relative entropy) as the metric. Suppose we have the normalized luminosity function from mock $n_{\rm model}(L)$ and the normalized luminosity function from ELG observation $n_{\rm ELG}(L)$, the KL-divergence of $n_{\rm model}(L)$ from $n_{\rm ELG}(L)$ is defined as:

\begin{equation}
    D_{\rm KL}(n_{\rm ELG} | n_{\rm mock}) = \int n_{\rm ELG}(L) \log\left(\frac{n_{\rm ELG}(L)}{n_{\rm mock}(L)}\right) d(L).
\end{equation}
Note that the luminosity function is normalized here, i.e., 
$$\int n_{\rm ELG}(L) d(L) = 1, ~~ \int n_{\rm model}(L) d(L) = 1.$$
The KL-divergence measures the information gain when using $n_{\rm model}(L)$ to approximate $n_{\rm ELG}(L)$. With smaller KL-divergence, the mismatch between $n_{\rm ELG}(L)$ and $n_{\rm model}(L)$ is smaller. The KL-divergence is always non-negative, with $D_{\rm KL}(n_{\rm ELG} | n_{\rm model}) = 0$ if and only if $n_{\rm ELG}(L) = n_{\rm model}(L)$ everywhere. 

We forbid parameter sets that yield too much discrepancy on $L[\rm OII]$ between model and ELG data by manually restricting $D_{\rm KL}(n_{\rm ELG} | n_{\rm mock}) < 0.01$. One can interpret this constraint as if we have a prior probability distribution $n_{\rm model}$ and we observe a posterior probability distribution $n_{\rm ELG}$, then the log-likelihood must be greater than -0.01. The reason we use this approach instead of using $\chi^2$ to constrain on $L[\rm OII]$ is that we have a relatively more accurate measurement on the $[\rm OII]$ luminosity than the two-point clustering. So if we combine the $\chi^2$ of $w_p$ with the $\chi^2$ of $L[\rm OII]$ as the full log-likelihood in the MCMC, the posterior distribution of parameters will be dominated by matching the luminosity function, which is not what we intended to do. Given the reason above, we choose a reasonable hard threshold cut for constraining the luminosity function and let the MCMC focus on matching the two-point correlation function.

In order to reproduce the small scale clustering of the eBOSS ELG sample, we constrain parameters using the projected correlation function $w_p(r_p)$. Let $\xi_{\rm model}$ denote the projected correlation function from our model given a given set of parameters and $\xi_{\rm ELG}$ denote the projected correlation function from eBOSS ELG sample. We utilize the covariance matrix $C(\xi_i, \xi_j)$ from GLAM-QPM mocks to describe the distance between $\xi_{\rm model}$ and $\xi_{\rm ELG}$. Assuming the $\xi_{\rm model}$ has a multivariate normal distribution around its expected value, the log-likelihood of obtaining $\xi_{\rm model}$ is:

\begin{equation}
    \log P(\xi_{\rm model}) = -\frac{1}{2}\log((2\pi)^k|C|)-\frac{1}{2}(\xi_{\rm model}-\xi_{\rm ELG})^TC^{-1}(\xi_{\rm model}-\xi_{\rm ELG}),
\end{equation}
where $k$ is the dimension of $\xi_{\rm model}$, in our case $k=17$. This assumption can be validated through 2000 GLAM-QPM mocks, we measure cross-$\chi^2$ with 12 data points between $0.34h^{-1}\textrm{Mpc}$ and $70h^{-1}\textrm{Mpc}$ and the results indeed follow $\chi^2$ distribution with 12 degrees of freedom.  Since $k$ and the covariance matrix $C$ are constants, the first term of $\log P(\xi_{\rm model})$ is a constant and doesn't change if our model parameters change.

\section{Results}\label{sec: result}
\subsection{ELG luminosity function and clustering measurements}
\begin{figure}
\begin{center}
\includegraphics[width=0.5\textwidth]{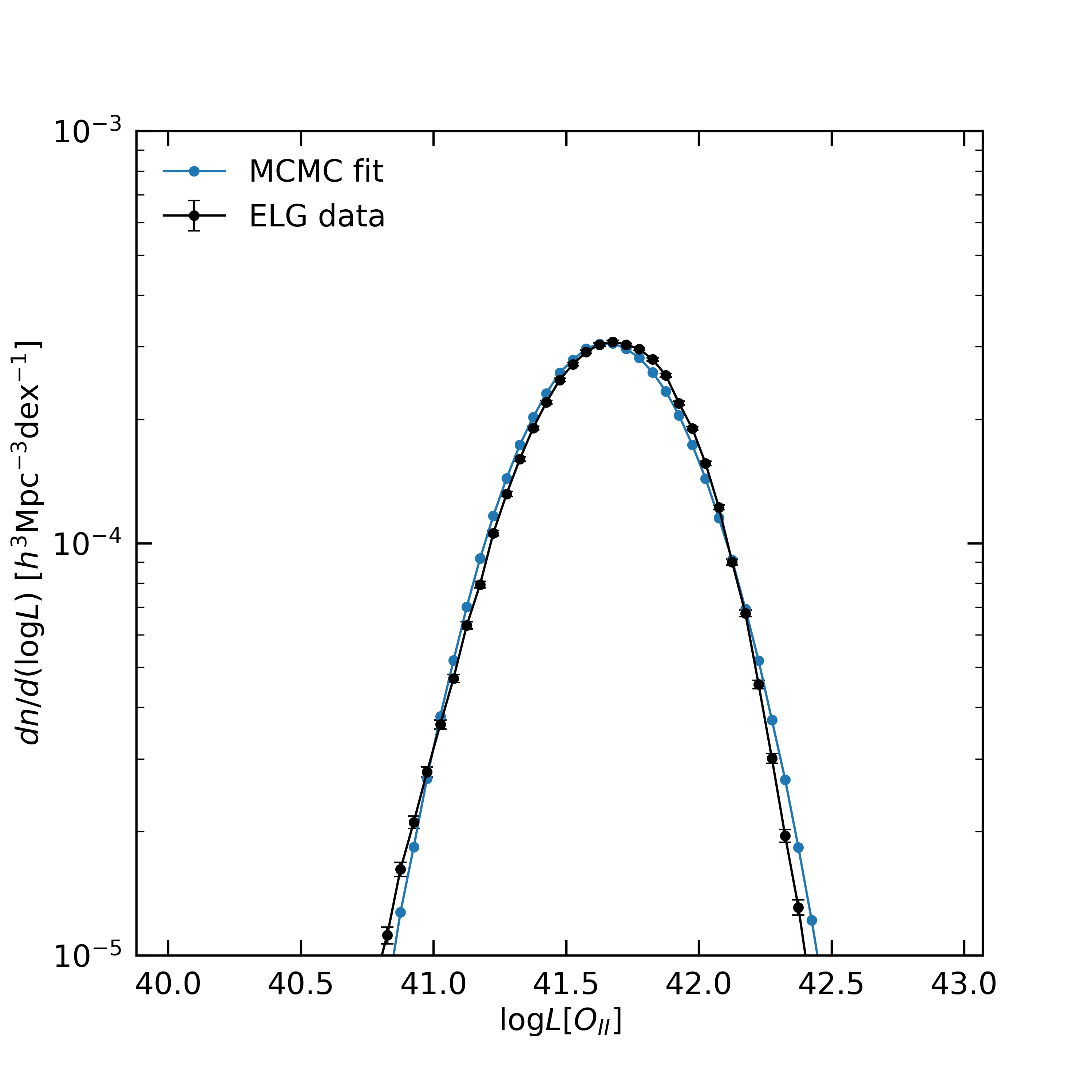}
\caption{The $[\rm OII]$ luminosity function of the eBOSS ELG sample and abundance matching fit. The black curve shows the $[\rm OII]$ luminosity function measured from the eBOSS ELG sample, using equally spaced binning of 0.05 dex in $\log L[\rm OII]$. The errorbars are estimated using bootstrap resampling method. The blue curve shows the model predicted $[\rm OII]$ luminosity function from the best fit parameters of MCMC, using the same binning as the black curve.}
\label{fig: LOII}
\end{center}
\end{figure}

The main goal of this work is to construct a model for the one-point and two-point statistics of the eBOSS ELG sample with conditional abundance matching. We run 600,000 MCMC iterations to constrain the three free parameters described in \S \ref{sec: method}. In Fig. \ref{fig: LOII} we present the comparison between the $[\rm OII]$ luminosity function of the eBOSS ELG sample and that from our abundance matching model. Within the space of models that pass the constraint of $D_{\rm KL}(n_{\rm ELG} | n_{\rm mock})<0.01$, we find a good match to the $w_p(r_p)$ data. Note that although the discrepancy between the ELG luminosity function and our model is small, the errorbars of the ELG luminosity function, estimated using bootstrap method, are even smaller. This illustrates the point we made in \S \ref{subsec: MCMC} that if we add $\chi^2$ of $L[\rm OII]$ in MCMC to optimize the parameters, the model will mainly focus on matching $L[\rm OII]$ instead of $w_p$. So our approach focuses on the two-point clustering instead but also has a good constraint on the one-point statistics.

\begin{figure}
\begin{center}
\includegraphics[width=0.5\textwidth]{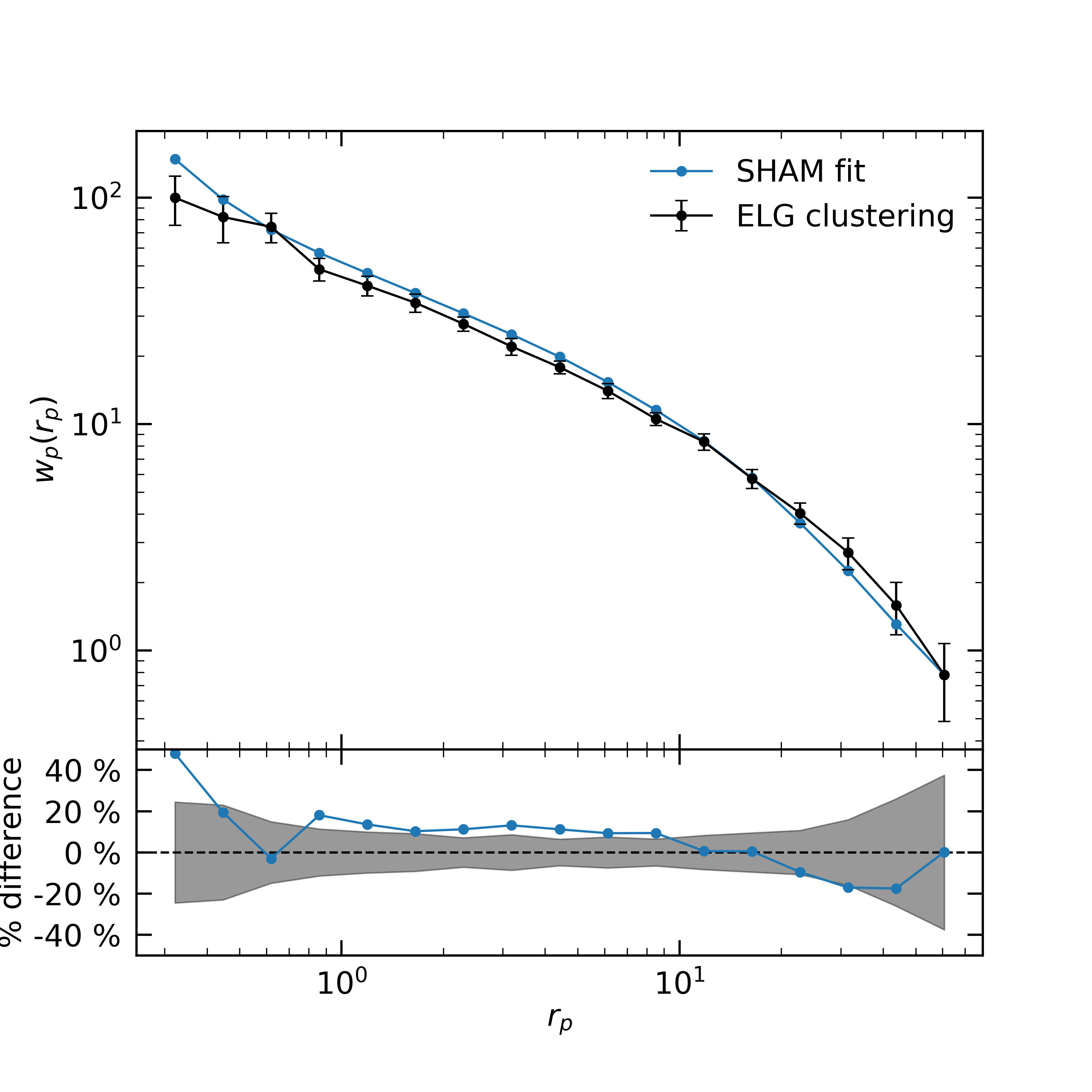}
\caption{The projected two-point correlation function $w_p(r_p)$ of the eBOSS ELG sample and our model result. In the top panel, the black curve shows the $w_p$ for the eBOSS ELG sample, the errorbars are estimated using jackknife resampling method with 25 jackknife bins. And the blue curve shows our model predicted $w_p$. The bottom panel shows the percentage difference between the data and the model, calculated as $(w_{p, \textrm{model}} - w_{p, \textrm{data}})/w_{p, \textrm{data}}$. The shaded area indicates the percentage of errorbars, calculated as $\pm e_{\textrm{data}} / w_{p, \textrm{data}}$.}
\label{fig: wp}
\end{center}
\end{figure}

In Fig. \ref{fig: wp} we show comparison of projected two-point correlation function between the eBOSS ELG sample and our model result, with the shaded regions as the 1$\sigma$ error. Overall the model and the ELG $w_p(r_p)$ agree well at the 1$\sigma$ level. The best fit of $\chi^2$ is $10.0$ with 17 degrees of freedom for $w_p$. The corresponding p-value is not significant at $p<0.05$, indicating that there is no statistical significant between our fitted model and the observed data. The reduced $\chi^2$ is slightly below 1, which is likely due to the method that we up-scale the errorbar of $\xi_{\textnormal{ELG}}$ in Sec. \ref{subsec: ssc}. At scales of $r_p\gtrsim 10 h^{-1}\rm Mpc$ our model produces relatively lower clustering, and at $r \lesssim 10 h^{-1}\rm Mpc$ our $w_p$ is higher than the eBOSS ELG sample. This is likely due to the reason that our quenched fraction approach yields a slightly more satellites than the eBOSS ELG sample, so our model predicted $w_p$ is higher at the transition scale between 1-halo term and 2-halo term.

\begin{figure*}
\begin{center}
\includegraphics[width=0.95\textwidth]{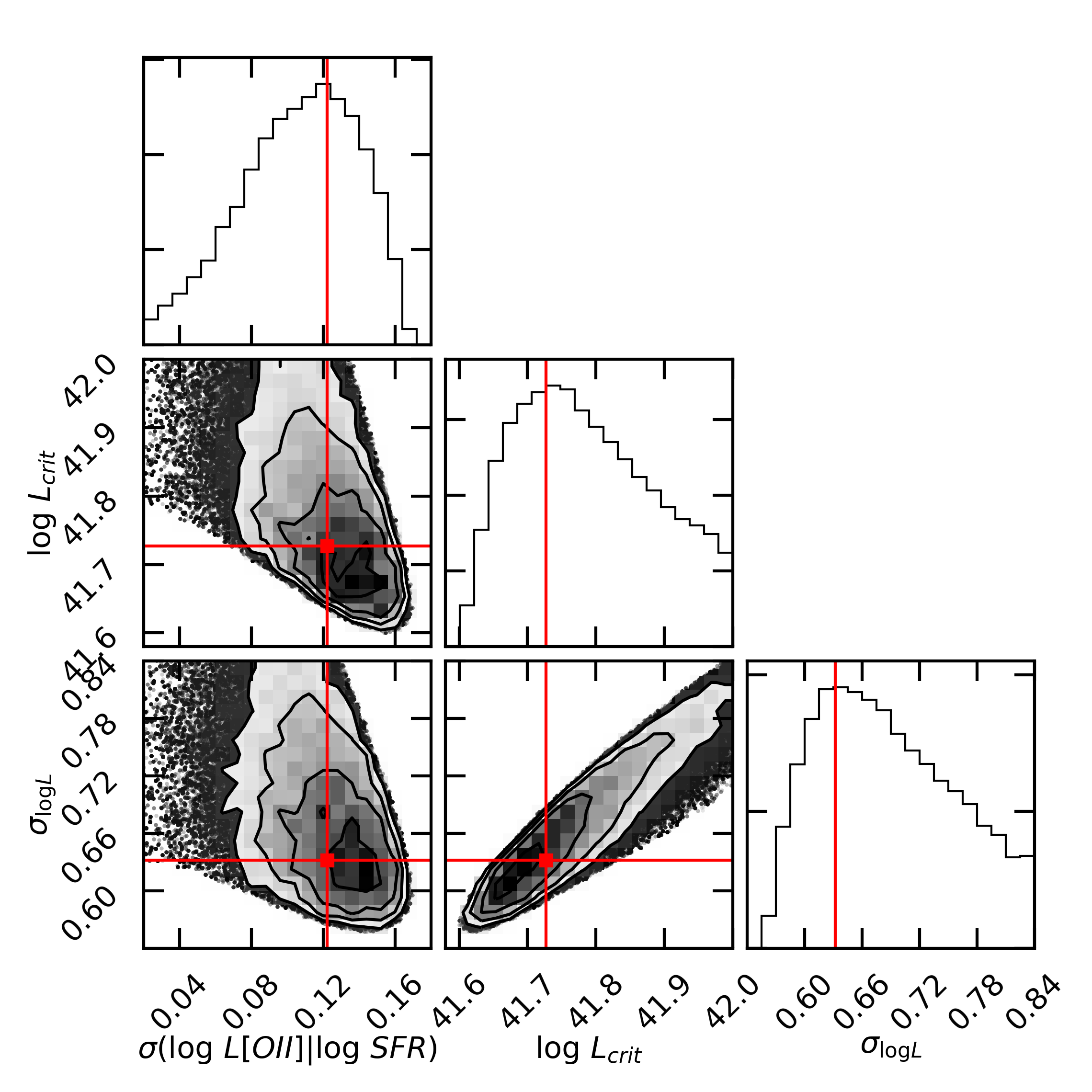}
\caption{The contours of likelihood of three parameters in our model, shown at 0.5, 1, 1.5 and 2 sigma. The diagnal panels represent the one-dimensional pdf of each parameter. The red points and red lines indicate the best fit parameters.}
\label{mcmc}
\end{center}
\end{figure*}

The constraints on the parameters are presented in Fig.~\ref{mcmc}. We find that the scatter in $\log L[\rm OII]$ at fixed SFR is $\sigma(L[\rm O_{II}] | SFR) = 0.11_{-0.04}^{+0.03}$ dex. While most of the literature focus on estimating SFR using [OII] flux, there is limited discussion of the scatter in $[\rm OII]$ luminosity. However we could utilize the results from \citet{Moustakas2006}, where they find that $\sigma(\textnormal{SFR}|L[\textnormal{OII}]) \sim 0.3$ dex for various dust corrections. Since the $L[\rm OII]$ is proportional to SFR according to Eq. \ref{eq: LOII}, and the empirical correction term is negatively correlated with SFR, the slope of $\log L[\rm OII]$ against SFR should be less than 1. Thus the $\sigma(L[\rm OII] | SFR)$ should be less than $\sigma(\rm SFR|L[OII]) \sim 0.3$, which justifies our finding.

We find that the incompleteness coefficient in Eq. \ref{eq: erf} is $\eta = 0.12$ under the best MCMC fit. This means that the eBOSS ELG sample selects 12\% of star-forming galaxies at mass range $10^{10}h^{-1}M_\odot \sim 10^{11.5}h^{-1}M_\odot$ from our mock catalog. This is in accordance with the HOD results of \citet{Guo2019}, where they find that the eBOSS ELG target selection only select 1-10\% of the star-forming galaxies at high mass end for various redshifts. \citet{Guo2019} also find that the completeness decreases at the low mass end. Accordingly, we use an error function to mimic the target selection at the low mass end and assume a universal incompleteness rate. Our best fit $\sigma_{\log L}=0.63$ dex, which model the sharpness of the $[\rm OII]$ selection due to the ELG target selection at the low $[\rm OII]$ end.

\subsection{Effect of galaxy assembly bias}
\begin{figure*}
\begin{center}
\includegraphics[width=0.95\textwidth]{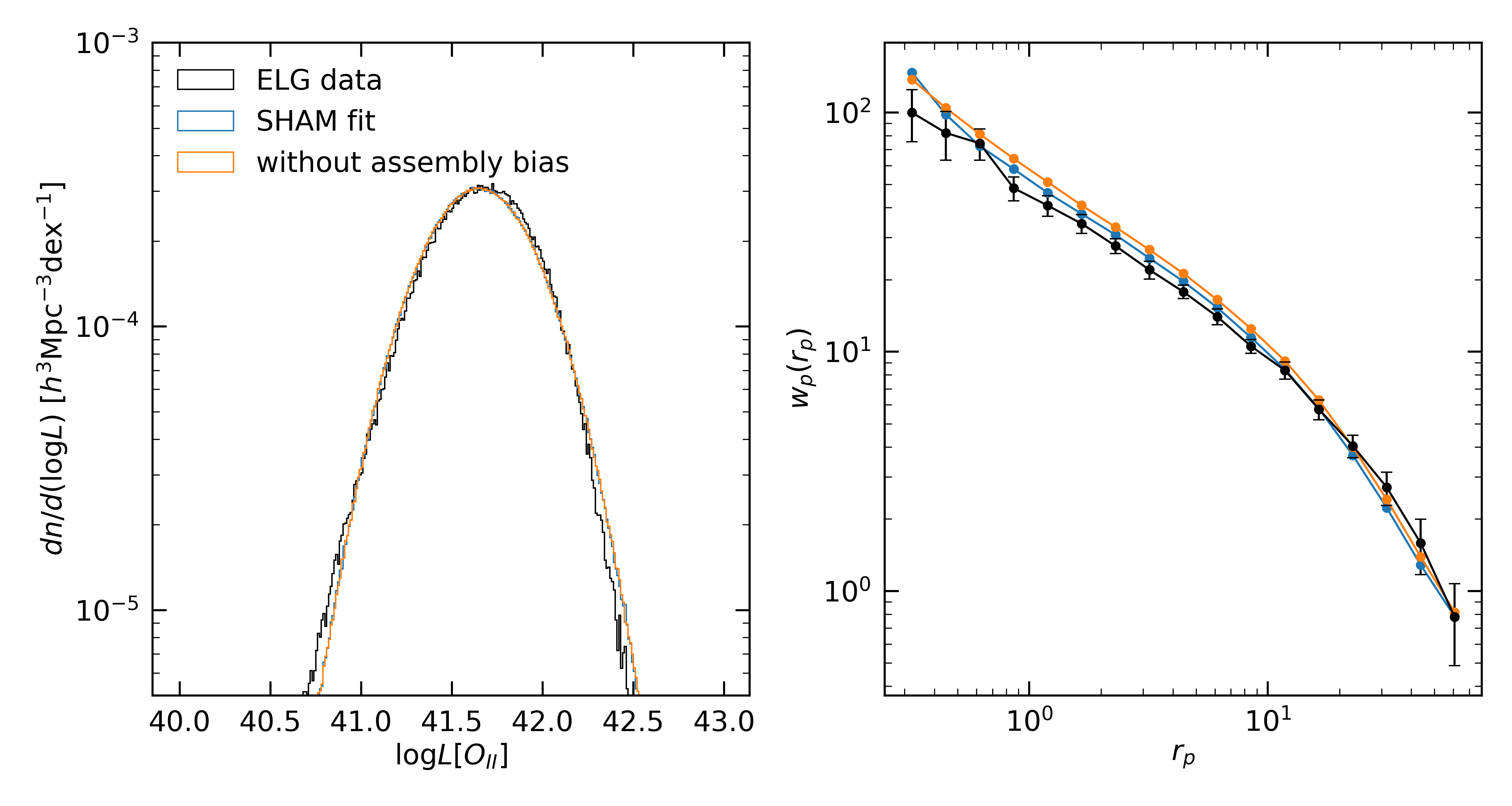}
\caption{The comparison between models with and without galaxy assembly bias. The left panel shows the $L[\rm OII]$ luminosity function and the right panel shows the projected two-point correlation function. The black lines show the observation from the eBOSS ELG sample, the blue lines show the model result with galaxy assembly bias, and the orange lines show the model results without galaxy assembly bias. The models are using the same methodology and the same parameter set, the only difference is whether we rank order SFR of galaxies with halo accretion rate. }
\label{fig: assembly bias}
\end{center}
\end{figure*}

In our conditional abundance matching approach we connect the SFR of galaxies with the accretion rate of halos in order to model the galaxy assembly bias. Here we define galaxy assembly bias as the dependence of galaxy properties on a second halo property that is related with halo assembly history. To test the impact of assembly bias, we perform a parallel analysis that repeats the fiducial model, with the exception of removing any correlation between SFR and halo growth rate. In that case, SFR only depends on stellar mass. In Fig. \ref{fig: assembly bias} we show the comparison between our best fit model and the model without the assembly bias. The two models have the same number density and satellite fraction. From the left panel in Fig. \ref{fig: assembly bias} we can see that both two models can match the $[\rm OII]$ luminosity well, and from the right panel we find that the two-point correlation function without assembly bias is generally $\sim 20\%$ higher than the model with assembly bias. By measuring the $\chi^2$ of the model without assembly bias, we find that removing the assembly bias causes 2PCF to deviate from the best fit and fits worse than our fiducial model. This result is expected: if the assembly bias is not included in our model, the star-forming galaxies will more likely reside in older halos given a fixed halo mass bin from $10^{11} h^{-1}M_\odot$ to $10^{14} h^{-1}M_\odot$. Thus the two-point clustering of galaxies will be higher because older halos are more clustered. 

\begin{figure}
\begin{center}
\includegraphics[width=0.5\textwidth]{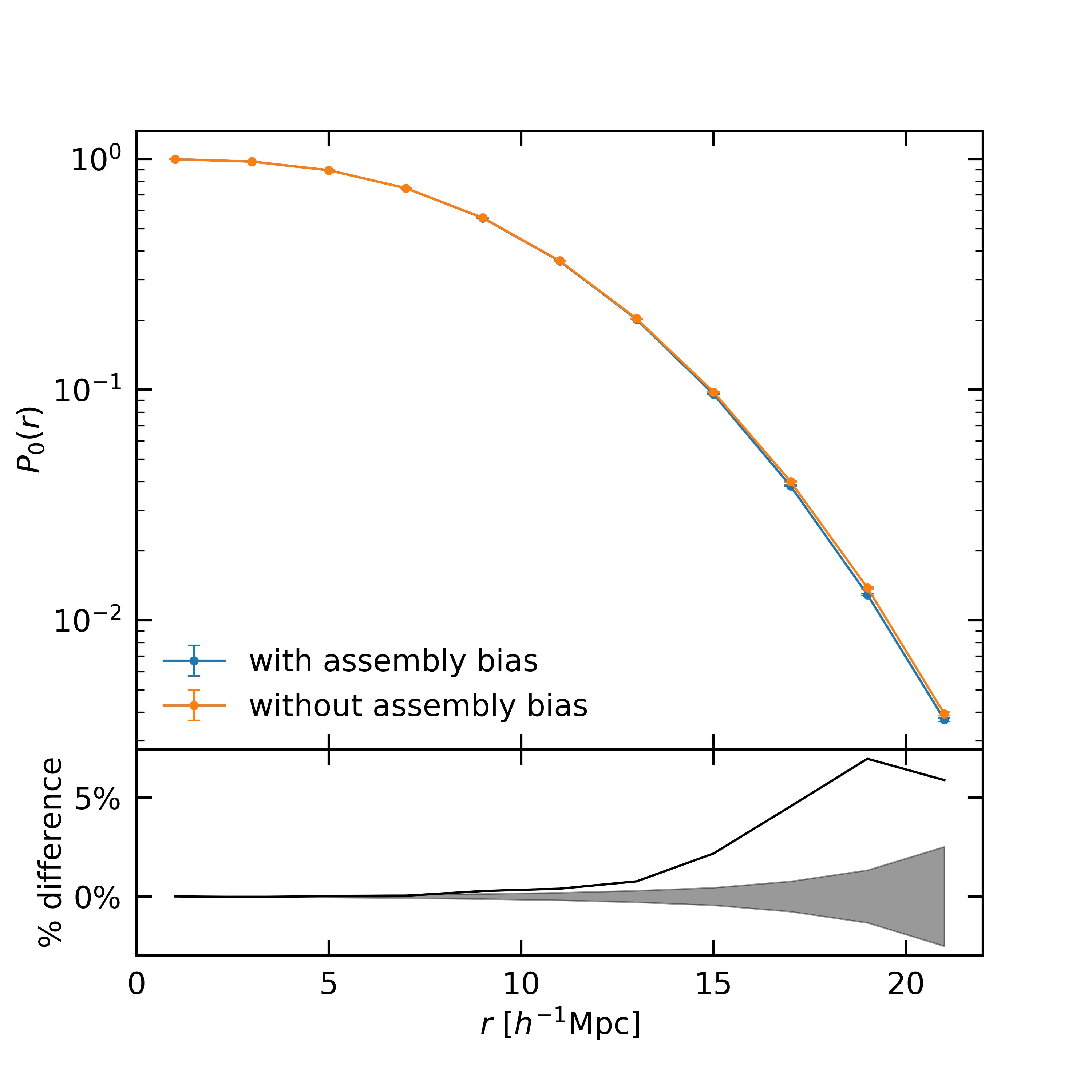}
\caption{The comparison of VPF between models with and without galaxy assembly bias. The upper panel shows the VPF, the blue line represents the model with assembly bias and the orange line shows the model without assembly bias. In the bottom panel, the black line shows the percentage difference between two models, and the grey area represents the errorbar of the difference. }
\label{fig: VPF}
\end{center}
\end{figure}

We also test the impact of galaxy assembly bias on the void probability function (VPF). The VPF $P_0(r)$ is defined as the probability of a randomly placed sphere with radius $r$ containing no galaxy of a certain type. Compared with the two-point statistics, it is a useful tool to test whether halo occupation or galaxy bias changes from high to low density environment \citep{tinker2006}. We make measurements for VPF in the range of $r\sim[1, 21] h^{-1}\rm Mpc$ with linear binning of $\Delta r=2 h^{-1}\rm Mpc$. For each radius we place $10^6$ spheres randomly in our mock catalog to obtain a relatively precise measurement of VPF. The errors on VPF are estimated using bootstrap method with 1000 resampling size. In Fig. \ref{fig: VPF} we show the difference of VPF with and without galaxy assembly bias. The model without assembly bias produces around 5\% larger VPF at scale of $r\sim 20 h^{-1}\rm Mpc$, as shown in the lower panel of Fig. \ref{fig: VPF}. Note that the difference is more than twice of the errorbar, meaning that it is statistically significant, and can be used as a test with current and upcoming data.

\subsection{Halo Occupation}

\begin{figure}
\begin{center}
\includegraphics[width=0.5\textwidth]{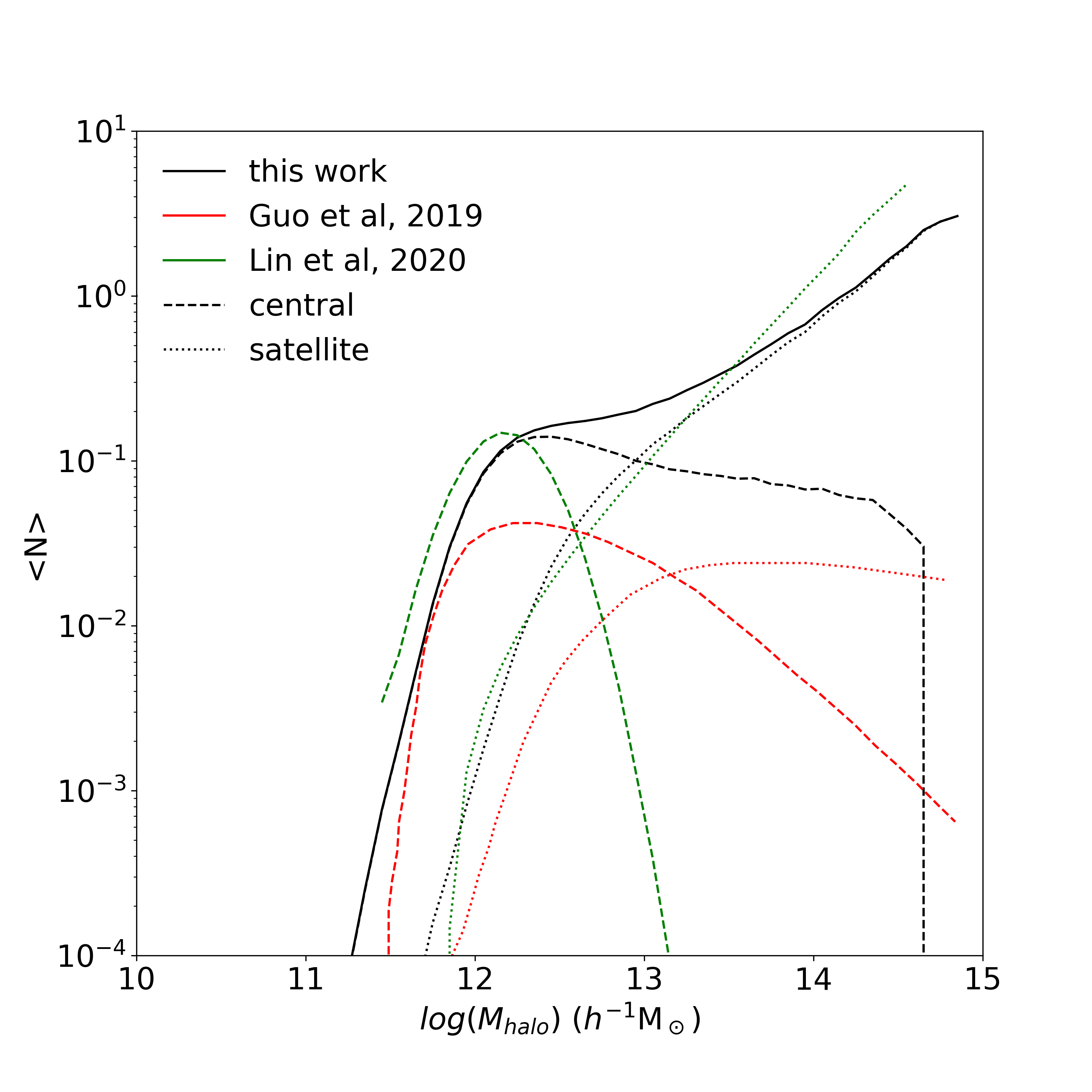}
\caption{Halo occupation distribution of our abundance matching model, compared with \citet{Guo2019} and \citet{Lin2020}. The black lines show our result, the red lines show the result from \citet{Guo2019} and the green lines show the result from \citet{Lin2020}. For all results, the dashed lines represent the HOD for central galaxies and the dotted lines represent the HOD for satellite galaxies.}
\label{ham_hod}
\end{center}
\end{figure}

We investigate the mean halo occupation function of our abundance matching model, as shown in Fig. \ref{ham_hod}. The centrals and the satellites are shown in black dashed line and black dotted line, respectively. The result indicates that ELG central galaxy can be modeled as a log-normal distribution at low mass end and a decreasing power-law at high mass end. According to our model, the occupation numbers of ELG at high halo mass end decreases, but not dramatically. We compare our result with \citet{Guo2019} at redshift $0.8<z<0.9$, as shown in the red line. Our amplitude of HOD function is higher than \citet{Guo2019} because we only show their results for redshift bin $0.8<z<0.9$. \citet{Guo2019} finds a significant decrease of the occupation function for centrals at high mass end, likely because they use a higher quenched fraction than we do. We also compare our result with \citet{Lin2020}, and we find a similar satellite HOD. The difference in the central HOD is due to \citet{Lin2020} adopting a Gaussian function for central galaxies.

Recently, \citet{Avila2020} explored different forms of HOD model to study the eBOSS ELG sample. They have tested three different formulations for the central HOD: a step-wise erf function, a Gaussian function, and a Gaussian + decaying power-law at high mass end. The difference between those three formulas is mainly at the high mass end. With the same power-law for the satellite HOD, they find that the shape of the central HOD has a less significant impact on the two-point clustering than the satellite fraction. The data marginally rules out the Gaussian formula, and there is negligible difference between the step-wise erf function and the Gaussian + decaying power-law, which is in accordance with our result.

According to our best fit model, the satellite fraction is $19.30\pm 0.02$\%, agreeing with an overall $\sim 20$ percent satellite fraction for all star-forming galaxies at $z=0.88$ \citep{Tinker2013}. \citet{Favole2016} also found that the satellite fraction of ELGs at redshift 0.8 is $22.5 \pm 2.5\%$. 

\section{Summary}\label{sec: conclusion}
This paper presents a methodology for constructing subhalo-based models of emission line galaxies that are physically motivated but contain limited number of free parameters that can be constrained by two-point statistics. Our method utilizes conditional abundance matching to model the one-point and two-point statistics of the eBOSS ELG sample. We create a mock galaxy catalog by matching the stellar mass of galaxies with the peak circular velocity of halos. We use quenched fraction from \citet{Tinker2013} to select star-forming galaxies. The eBOSS ELG target selection is modeled as an error function of $[\rm OII]$ luminosity. We have demonstrated that by conditionally matching the SFR of galaxies and halo growth rate, we are able to model the one-point and two-point statistics as well as the assembly bias of the eBOSS ELG sample. The free parameters in our model are all related to $[\rm OII]$ luminosity, and are determined by the MCMC method. 

Our model is able to reproduce the projected two-point correlation function of eBOSS ELG within 1$\sigma$ error. According to our model, we find that the eBOSS ELG sample only selects $\sim 12\%$ of the star-forming galaxies at $10^{10} h^{-1}M_\odot \lesssim M_\star \lesssim 10^{11.5} h^{-1}M_\odot$, agreeing with the findings in \citet{Guo2019}. The satellite fraction is 19.4\%, which is very similar to the $\sim 20\%$ satellite fraction for star-forming galaxies in \citet{Tinker2013}. 

The star formation activity at $z\sim 0-2$ makes ELGs ideal tracers for spectroscopic surveys. The approach that we develop in this paper can be used for testing galaxy formation physics, as well as producing high-fidelity mock galaxy catalogs for large-scale structure analysis. Although we present an analysis of eBOSS data, the method is flexible and can be applied to any survey with an ELG-type sample selection.

\section*{acknowledgments}

SL is grateful to support from the CCPP at New York University. SL, JLT and MRB are supported by NSF Award 1615997.

Funding for the Sloan Digital Sky Survey IV has been provided by
the Alfred P. Sloan Foundation, the U.S. Department of Energy Office of
Science, and the Participating Institutions. SDSS-IV acknowledges
support and resources from the Center for High-Performance Computing at
the University of Utah. The SDSS web site is www.sdss.org.

SDSS-IV is managed by the Astrophysical Research Consortium for the 
Participating Institutions of the SDSS Collaboration including the 
Brazilian Participation Group, the Carnegie Institution for Science, 
Carnegie Mellon University, the Chilean Participation Group, the French Participation Group, Harvard-Smithsonian Center for Astrophysics, 
Instituto de Astrof\'isica de Canarias, The Johns Hopkins University, 
Kavli Institute for the Physics and Mathematics of the Universe (IPMU) / 
University of Tokyo, Lawrence Berkeley National Laboratory, 
Leibniz Institut f\"ur Astrophysik Potsdam (AIP),  
Max-Planck-Institut f\"ur Astronomie (MPIA Heidelberg), 
Max-Planck-Institut f\"ur Astrophysik (MPA Garching), 
Max-Planck-Institut f\"ur Extraterrestrische Physik (MPE), 
National Astronomical Observatory of China, New Mexico State University, 
New York University, University of Notre Dame, 
Observat\'ario Nacional / MCTI, The Ohio State University, 
Pennsylvania State University, Shanghai Astronomical Observatory, 
United Kingdom Participation Group,
Universidad Nacional Aut\'onoma de M\'exico, University of Arizona, 
University of Colorado Boulder, University of Oxford, University of Portsmouth, 
University of Utah, University of Virginia, University of Washington, University of Wisconsin, 
Vanderbilt University, and Yale University.

\section*{Data availability}
The eBOSS ELG catalogs and the GLAM-QPM mock galaxy catalogs are available from the eBOSS DR16 galaxy catalog release.

\bibliographystyle{mnras}
\bibliography{main}

\begin{thebibliography}{}
\makeatletter
\relax
\def\mn@urlcharsother{\let\do\@makeother \do\$\do\&\do\#\do\^\do\_\do\%\do\~}
\def\mn@doi{\begingroup\mn@urlcharsother \@ifnextchar [ {\mn@doi@}
  {\mn@doi@[]}}
\def\mn@doi@[#1]#2{\def\@tempa{#1}\ifx\@tempa\@empty \href
  {http://dx.doi.org/#2} {doi:#2}\else \href {http://dx.doi.org/#2} {#1}\fi
  \endgroup}
\def\mn@eprint#1#2{\mn@eprint@#1:#2::\@nil}
\def\mn@eprint@arXiv#1{\href {http://arxiv.org/abs/#1} {{\tt arXiv:#1}}}
\def\mn@eprint@dblp#1{\href {http://dblp.uni-trier.de/rec/bibtex/#1.xml}
  {dblp:#1}}
\def\mn@eprint@#1:#2:#3:#4\@nil{\def\@tempa {#1}\def\@tempb {#2}\def\@tempc
  {#3}\ifx \@tempc \@empty \let \@tempc \@tempb \let \@tempb \@tempa \fi \ifx
  \@tempb \@empty \def\@tempb {arXiv}\fi \@ifundefined
  {mn@eprint@\@tempb}{\@tempb:\@tempc}{\expandafter \expandafter \csname
  mn@eprint@\@tempb\endcsname \expandafter{\@tempc}}}

\bibitem[\protect\citeauthoryear{{Alam} et~al.,}{{Alam}
  et~al.}{2021}]{eboss2020}
{Alam} S.,  et~al., 2021, \mn@doi [\prd] {10.1103/PhysRevD.103.083533}, \href
  {https://ui.adsabs.harvard.edu/abs/2021PhRvD.103h3533A} {103, 083533}

\bibitem[\protect\citeauthoryear{{Avila} et~al.,}{{Avila}
  et~al.}{2020}]{Avila2020}
{Avila} S.,  et~al., 2020, \mn@doi [\mnras] {10.1093/mnras/staa2951}, \href
  {https://ui.adsabs.harvard.edu/abs/2020MNRAS.499.5486A} {499, 5486}

\bibitem[\protect\citeauthoryear{{Behroozi}, {Conroy}  \&
  {Wechsler}}{{Behroozi} et~al.}{2010}]{behroozi2010}
{Behroozi} P.~S.,  {Conroy} C.,   {Wechsler} R.~H.,  2010, \mn@doi [\apj]
  {10.1088/0004-637X/717/1/379}, \href
  {http://adsabs.harvard.edu/abs/2010ApJ...717..379B} {717, 379}

\bibitem[\protect\citeauthoryear{{Behroozi}, {Wechsler}  \& {Wu}}{{Behroozi}
  et~al.}{2013}]{Behroozi2013}
{Behroozi} P.~S.,  {Wechsler} R.~H.,   {Wu} H.-Y.,  2013, \mn@doi [\apj]
  {10.1088/0004-637X/762/2/109}, \href
  {https://ui.adsabs.harvard.edu/abs/2013ApJ...762..109B} {762, 109}

\bibitem[\protect\citeauthoryear{{Benson}, {Cole}, {Frenk}, {Baugh}  \&
  {Lacey}}{{Benson} et~al.}{2000}]{Benson2000}
{Benson} A.~J.,  {Cole} S.,  {Frenk} C.~S.,  {Baugh} C.~M.,   {Lacey} C.~G.,
  2000, \mn@doi [\mnras] {10.1046/j.1365-8711.2000.03101.x}, \href
  {http://adsabs.harvard.edu/abs/2000MNRAS.311..793B} {311, 793}

\bibitem[\protect\citeauthoryear{{Blanton} et~al.,}{{Blanton}
  et~al.}{2017}]{blanton2017}
{Blanton} M.~R.,  et~al., 2017, \mn@doi [\aj] {10.3847/1538-3881/aa7567}, \href
  {https://ui.adsabs.harvard.edu/abs/2017AJ....154...28B} {154, 28}

\bibitem[\protect\citeauthoryear{{Cao}, {Tinker}, {Mao}  \& {Wechsler}}{{Cao}
  et~al.}{2020}]{Cao2020}
{Cao} J.-z.,  {Tinker} J.~L.,  {Mao} Y.-Y.,   {Wechsler} R.~H.,  2020, \mn@doi
  [\mnras] {10.1093/mnras/staa2644}, \href
  {https://ui.adsabs.harvard.edu/abs/2020MNRAS.498.5080C} {498, 5080}

\bibitem[\protect\citeauthoryear{{Chaves-Montero}, {Angulo}, {Schaye},
  {Schaller}, {Crain}, {Furlong}  \& {Theuns}}{{Chaves-Montero}
  et~al.}{2016}]{Chaves2016}
{Chaves-Montero} J.,  {Angulo} R.~E.,  {Schaye} J.,  {Schaller} M.,  {Crain}
  R.~A.,  {Furlong} M.,   {Theuns} T.,  2016, \mn@doi [\mnras]
  {10.1093/mnras/stw1225}, \href
  {https://ui.adsabs.harvard.edu/abs/2016MNRAS.460.3100C} {460, 3100}

\bibitem[\protect\citeauthoryear{{Cole}, {Lacey}, {Baugh}  \& {Frenk}}{{Cole}
  et~al.}{2000}]{Cole2000}
{Cole} S.,  {Lacey} C.~G.,  {Baugh} C.~M.,   {Frenk} C.~S.,  2000, \mn@doi
  [\mnras] {10.1046/j.1365-8711.2000.03879.x}, \href
  {https://ui.adsabs.harvard.edu/abs/2000MNRAS.319..168C} {319, 168}

\bibitem[\protect\citeauthoryear{{Conroy}, {Wechsler}  \& {Kravtsov}}{{Conroy}
  et~al.}{2006}]{Conroy2006}
{Conroy} C.,  {Wechsler} R.~H.,   {Kravtsov} A.~V.,  2006, \mn@doi [\apj]
  {10.1086/503602}, \href
  {https://ui.adsabs.harvard.edu/abs/2006ApJ...647..201C} {647, 201}

\bibitem[\protect\citeauthoryear{{DESI Collaboration} et~al.,}{{DESI
  Collaboration} et~al.}{2016a}]{DESI2016a}
{DESI Collaboration} et~al., 2016a, arXiv e-prints, \href
  {https://ui.adsabs.harvard.edu/abs/2016arXiv161100036D} {p. arXiv:1611.00036}

\bibitem[\protect\citeauthoryear{{DESI Collaboration} et~al.,}{{DESI
  Collaboration} et~al.}{2016b}]{DESI2016b}
{DESI Collaboration} et~al., 2016b, arXiv e-prints, \href
  {https://ui.adsabs.harvard.edu/abs/2016arXiv161100037D} {p. arXiv:1611.00037}

\bibitem[\protect\citeauthoryear{{Dawson} et~al.,}{{Dawson}
  et~al.}{2016}]{dawson2016}
{Dawson} K.~S.,  et~al., 2016, \mn@doi [\aj] {10.3847/0004-6256/151/2/44},
  \href {https://ui.adsabs.harvard.edu/abs/2016AJ....151...44D} {151, 44}

\bibitem[\protect\citeauthoryear{{Dey} et~al.,}{{Dey} et~al.}{2018}]{Dey2018}
{Dey} A.,  et~al., 2018, arXiv e-prints, \href
  {http://adsabs.harvard.edu/abs/2018arXiv180408657D} {}

\bibitem[\protect\citeauthoryear{{Favole} et~al.,}{{Favole}
  et~al.}{2016}]{Favole2016}
{Favole} G.,  et~al., 2016, \mn@doi [\mnras] {10.1093/mnras/stw1483}, \href
  {https://ui.adsabs.harvard.edu/abs/2016MNRAS.461.3421F} {461, 3421}

\bibitem[\protect\citeauthoryear{{Feldman}, {Kaiser}  \& {Peacock}}{{Feldman}
  et~al.}{1994}]{fkp1994}
{Feldman} H.~A.,  {Kaiser} N.,   {Peacock} J.~A.,  1994, \mn@doi [\apj]
  {10.1086/174036}, \href {http://adsabs.harvard.edu/abs/1994ApJ...426...23F}
  {426, 23}

\bibitem[\protect\citeauthoryear{{Foreman-Mackey}, {Hogg}, {Lang}  \&
  {Goodman}}{{Foreman-Mackey} et~al.}{2013}]{emcee2013}
{Foreman-Mackey} D.,  {Hogg} D.~W.,  {Lang} D.,   {Goodman} J.,  2013, \mn@doi
  [\pasp] {10.1086/670067}, \href
  {https://ui.adsabs.harvard.edu/abs/2013PASP..125..306F} {125, 306}

\bibitem[\protect\citeauthoryear{{Gilbank}, {Baldry}, {Balogh}, {Glazebrook}
  \& {Bower}}{{Gilbank} et~al.}{2010}]{Gilbank2010}
{Gilbank} D.~G.,  {Baldry} I.~K.,  {Balogh} M.~L.,  {Glazebrook} K.,   {Bower}
  R.~G.,  2010, \mn@doi [\mnras] {10.1111/j.1365-2966.2010.16640.x}, \href
  {https://ui.adsabs.harvard.edu/abs/2010MNRAS.405.2594G} {405, 2594}

\bibitem[\protect\citeauthoryear{{Goodman} \& {Weare}}{{Goodman} \&
  {Weare}}{2010}]{Goodman2010}
{Goodman} J.,  {Weare} J.,  2010, \mn@doi [Communications in Applied
  Mathematics and Computational Science] {10.2140/camcos.2010.5.65}, \href
  {https://ui.adsabs.harvard.edu/abs/2010CAMCS...5...65G} {5, 65}

\bibitem[\protect\citeauthoryear{{Gunn} et~al.,}{{Gunn}
  et~al.}{2006}]{Gunn2006}
{Gunn} J.~E.,  et~al., 2006, \mn@doi [\aj] {10.1086/500975}, \href
  {https://ui.adsabs.harvard.edu/abs/2006AJ....131.2332G} {131, 2332}

\bibitem[\protect\citeauthoryear{{Guo} et~al.,}{{Guo} et~al.}{2019}]{Guo2019}
{Guo} H.,  et~al., 2019, \mn@doi [\apj] {10.3847/1538-4357/aaf9ad}, \href
  {https://ui.adsabs.harvard.edu/abs/2019ApJ...871..147G} {871, 147}

\bibitem[\protect\citeauthoryear{{Hearin} \& {Watson}}{{Hearin} \&
  {Watson}}{2013}]{Hearin2013}
{Hearin} A.~P.,  {Watson} D.~F.,  2013, \mn@doi [\mnras]
  {10.1093/mnras/stt1374}, \href
  {https://ui.adsabs.harvard.edu/abs/2013MNRAS.435.1313H} {435, 1313}

\bibitem[\protect\citeauthoryear{{Hearin}, {Watson}, {Becker}, {Reyes},
  {Berlind}  \& {Zentner}}{{Hearin} et~al.}{2014}]{Hearin2014}
{Hearin} A.~P.,  {Watson} D.~F.,  {Becker} M.~R.,  {Reyes} R.,  {Berlind}
  A.~A.,   {Zentner} A.~R.,  2014, \mn@doi [\mnras] {10.1093/mnras/stu1443},
  \href {https://ui.adsabs.harvard.edu/abs/2014MNRAS.444..729H} {444, 729}

\bibitem[\protect\citeauthoryear{{Hearin}, {Zentner}, {van den Bosch},
  {Campbell}  \& {Tollerud}}{{Hearin} et~al.}{2016}]{Hearin2016}
{Hearin} A.~P.,  {Zentner} A.~R.,  {van den Bosch} F.~C.,  {Campbell} D.,
  {Tollerud} E.,  2016, \mn@doi [\mnras] {10.1093/mnras/stw840}, \href
  {https://ui.adsabs.harvard.edu/abs/2016MNRAS.460.2552H} {460, 2552}

\bibitem[\protect\citeauthoryear{{Hearin}, {Behroozi}, {Kravtsov}  \&
  {Moster}}{{Hearin} et~al.}{2017}]{Hearin2017}
{Hearin} A.,  {Behroozi} P.,  {Kravtsov} A.,   {Moster} B.,  2017, arXiv
  e-prints, \href {https://ui.adsabs.harvard.edu/abs/2017arXiv171110500H} {p.
  arXiv:1711.10500}

\bibitem[\protect\citeauthoryear{{Ilbert} et~al.,}{{Ilbert}
  et~al.}{2013}]{ilbert2013}
{Ilbert} O.,  et~al., 2013, \mn@doi [\aap] {10.1051/0004-6361/201321100}, \href
  {http://adsabs.harvard.edu/abs/2013A%26A...556A..55I} {556, A55}

\bibitem[\protect\citeauthoryear{{Kauffmann}, {White}  \&
  {Guiderdoni}}{{Kauffmann} et~al.}{1993}]{Kauffmann1993}
{Kauffmann} G.,  {White} S.~D.~M.,   {Guiderdoni} B.,  1993, \mn@doi [\mnras]
  {10.1093/mnras/264.1.201}, \href
  {https://ui.adsabs.harvard.edu/abs/1993MNRAS.264..201K} {264, 201}

\bibitem[\protect\citeauthoryear{{Klypin}, {Yepes}, {Gottl{\"o}ber}, {Prada}
  \& {He{\ss}}}{{Klypin} et~al.}{2016}]{klypin2016}
{Klypin} A.,  {Yepes} G.,  {Gottl{\"o}ber} S.,  {Prada} F.,   {He{\ss}} S.,
  2016, \mn@doi [\mnras] {10.1093/mnras/stw248}, \href
  {https://ui.adsabs.harvard.edu/abs/2016MNRAS.457.4340K} {457, 4340}

\bibitem[\protect\citeauthoryear{{Kravtsov}, {Berlind}, {Wechsler}, {Klypin},
  {Gottl{\"o}ber}, {Allgood}  \& {Primack}}{{Kravtsov}
  et~al.}{2004}]{Kravtsov2004}
{Kravtsov} A.~V.,  {Berlind} A.~A.,  {Wechsler} R.~H.,  {Klypin} A.~A.,
  {Gottl{\"o}ber} S.,  {Allgood} B.,   {Primack} J.~R.,  2004, \mn@doi [\apj]
  {10.1086/420959}, \href
  {https://ui.adsabs.harvard.edu/abs/2004ApJ...609...35K} {609, 35}

\bibitem[\protect\citeauthoryear{{Lacey} \& {Cole}}{{Lacey} \&
  {Cole}}{1993}]{lacey1993}
{Lacey} C.,  {Cole} S.,  1993, \mn@doi [\mnras] {10.1093/mnras/262.3.627},
  \href {https://ui.adsabs.harvard.edu/abs/1993MNRAS.262..627L} {262, 627}

\bibitem[\protect\citeauthoryear{{Landy} \& {Szalay}}{{Landy} \&
  {Szalay}}{1993}]{Landy1993}
{Landy} S.~D.,  {Szalay} A.~S.,  1993, \mn@doi [\apj] {10.1086/172900}, \href
  {http://adsabs.harvard.edu/abs/1993ApJ...412...64L} {412, 64}

\bibitem[\protect\citeauthoryear{{Laureijs} et~al.,}{{Laureijs}
  et~al.}{2011}]{Laureijs2011}
{Laureijs} R.,  et~al., 2011, arXiv e-prints, \href
  {https://ui.adsabs.harvard.edu/abs/2011arXiv1110.3193L} {p. arXiv:1110.3193}

\bibitem[\protect\citeauthoryear{{Lee} et~al.,}{{Lee} et~al.}{2015}]{Lee2015}
{Lee} N.,  et~al., 2015, \mn@doi [\apj] {10.1088/0004-637X/801/2/80}, \href
  {http://adsabs.harvard.edu/abs/2015ApJ...801...80L} {801, 80}

\bibitem[\protect\citeauthoryear{{Lin} et~al.,}{{Lin} et~al.}{2020}]{Lin2020}
{Lin} S.,  et~al., 2020, \mn@doi [\mnras] {10.1093/mnras/staa2571}, \href
  {https://ui.adsabs.harvard.edu/abs/2020MNRAS.498.5251L} {498, 5251}

\bibitem[\protect\citeauthoryear{{Madau} \& {Dickinson}}{{Madau} \&
  {Dickinson}}{2014}]{Madau2014}
{Madau} P.,  {Dickinson} M.,  2014, \mn@doi [\araa]
  {10.1146/annurev-astro-081811-125615}, \href
  {https://ui.adsabs.harvard.edu/abs/2014ARA&A..52..415M} {52, 415}

\bibitem[\protect\citeauthoryear{{Moustakas}, {Kennicutt}  \&
  {Tremonti}}{{Moustakas} et~al.}{2006}]{Moustakas2006}
{Moustakas} J.,  {Kennicutt} Robert~C. J.,   {Tremonti} C.~A.,  2006, \mn@doi
  [\apj] {10.1086/500964}, \href
  {https://ui.adsabs.harvard.edu/abs/2006ApJ...642..775M} {642, 775}

\bibitem[\protect\citeauthoryear{{Noeske} et~al.,}{{Noeske}
  et~al.}{2007}]{Noeske2007}
{Noeske} K.~G.,  et~al., 2007, \mn@doi [\apjl] {10.1086/517927}, \href
  {https://ui.adsabs.harvard.edu/abs/2007ApJ...660L..47N} {660, L47}

\bibitem[\protect\citeauthoryear{{Peacock} \& {Smith}}{{Peacock} \&
  {Smith}}{2000}]{Peacock2000}
{Peacock} J.~A.,  {Smith} R.~E.,  2000, \mn@doi [\mnras]
  {10.1046/j.1365-8711.2000.03779.x}, \href
  {http://adsabs.harvard.edu/abs/2000MNRAS.318.1144P} {318, 1144}

\bibitem[\protect\citeauthoryear{{Peebles}}{{Peebles}}{1980}]{Peebles1980}
{Peebles} P.~J.~E.,  1980, {The large-scale structure of the universe}

\bibitem[\protect\citeauthoryear{{Prada}, {Klypin}, {Cuesta}, {Betancort-Rijo}
  \& {Primack}}{{Prada} et~al.}{2012}]{prada2012}
{Prada} F.,  {Klypin} A.~A.,  {Cuesta} A.~J.,  {Betancort-Rijo} J.~E.,
  {Primack} J.,  2012, \mn@doi [\mnras] {10.1111/j.1365-2966.2012.21007.x},
  \href {http://adsabs.harvard.edu/abs/2012MNRAS.423.3018P} {423, 3018}

\bibitem[\protect\citeauthoryear{{Raichoor} et~al.,}{{Raichoor}
  et~al.}{2017}]{raichoor2017}
{Raichoor} A.,  et~al., 2017, \mn@doi [\mnras] {10.1093/mnras/stx1790}, \href
  {http://adsabs.harvard.edu/abs/2017MNRAS.471.3955R} {471, 3955}

\bibitem[\protect\citeauthoryear{{Raichoor} et~al.,}{{Raichoor}
  et~al.}{2021}]{raichoor2021}
{Raichoor} A.,  et~al., 2021, \mn@doi [\mnras] {10.1093/mnras/staa3336}, \href
  {https://ui.adsabs.harvard.edu/abs/2021MNRAS.500.3254R} {500, 3254}

\bibitem[\protect\citeauthoryear{{Reddick}, {Wechsler}, {Tinker}  \&
  {Behroozi}}{{Reddick} et~al.}{2013}]{Reddick2013}
{Reddick} R.~M.,  {Wechsler} R.~H.,  {Tinker} J.~L.,   {Behroozi} P.~S.,  2013,
  \mn@doi [\apj] {10.1088/0004-637X/771/1/30}, \href
  {https://ui.adsabs.harvard.edu/abs/2013ApJ...771...30R} {771, 30}

\bibitem[\protect\citeauthoryear{{Rodr{\'\i}guez-Torres}
  et~al.,}{{Rodr{\'\i}guez-Torres} et~al.}{2016}]{Rodr2016}
{Rodr{\'\i}guez-Torres} S.~A.,  et~al., 2016, \mn@doi [\mnras]
  {10.1093/mnras/stw1014}, \href
  {https://ui.adsabs.harvard.edu/abs/2016MNRAS.460.1173R} {460, 1173}

\bibitem[\protect\citeauthoryear{{Schaye} et~al.,}{{Schaye}
  et~al.}{2015}]{Schaye2015}
{Schaye} J.,  et~al., 2015, \mn@doi [\mnras] {10.1093/mnras/stu2058}, \href
  {https://ui.adsabs.harvard.edu/abs/2015MNRAS.446..521S} {446, 521}

\bibitem[\protect\citeauthoryear{{Scoccimarro}, {Sheth}, {Hui}  \&
  {Jain}}{{Scoccimarro} et~al.}{2001}]{scoccimarro2001}
{Scoccimarro} R.,  {Sheth} R.~K.,  {Hui} L.,   {Jain} B.,  2001, \mn@doi [\apj]
  {10.1086/318261}, \href {http://adsabs.harvard.edu/abs/2001ApJ...546...20S}
  {546, 20}

\bibitem[\protect\citeauthoryear{{Seljak}}{{Seljak}}{2000}]{Seljak2000}
{Seljak} U.,  2000, \mn@doi [\mnras] {10.1046/j.1365-8711.2000.03715.x}, \href
  {http://adsabs.harvard.edu/abs/2000MNRAS.318..203S} {318, 203}

\bibitem[\protect\citeauthoryear{{Smee} et~al.,}{{Smee}
  et~al.}{2013}]{Smee2013}
{Smee} S.~A.,  et~al., 2013, \mn@doi [\aj] {10.1088/0004-6256/146/2/32}, \href
  {https://ui.adsabs.harvard.edu/abs/2013AJ....146...32S} {146, 32}

\bibitem[\protect\citeauthoryear{{Somerville} \& {Primack}}{{Somerville} \&
  {Primack}}{1999}]{Somerville1999}
{Somerville} R.~S.,  {Primack} J.~R.,  1999, \mn@doi [\mnras]
  {10.1046/j.1365-8711.1999.03032.x}, \href
  {https://ui.adsabs.harvard.edu/abs/1999MNRAS.310.1087S} {310, 1087}

\bibitem[\protect\citeauthoryear{{Spergel} et~al.,}{{Spergel}
  et~al.}{2013}]{Spergel2013}
{Spergel} D.,  et~al., 2013, arXiv e-prints, \href
  {https://ui.adsabs.harvard.edu/abs/2013arXiv1305.5422S} {p. arXiv:1305.5422}

\bibitem[\protect\citeauthoryear{{Springel}}{{Springel}}{2005}]{Springel2005}
{Springel} V.,  2005, \mn@doi [\mnras] {10.1111/j.1365-2966.2005.09655.x},
  \href {https://ui.adsabs.harvard.edu/abs/2005MNRAS.364.1105S} {364, 1105}

\bibitem[\protect\citeauthoryear{{Takada} et~al.,}{{Takada}
  et~al.}{2014}]{takada2014}
{Takada} M.,  et~al., 2014, \mn@doi [\pasj] {10.1093/pasj/pst019}, \href
  {https://ui.adsabs.harvard.edu/abs/2014PASJ...66R...1T} {66, R1}

\bibitem[\protect\citeauthoryear{{Taylor} et~al.,}{{Taylor}
  et~al.}{2020}]{Taylor2020}
{Taylor} E.~N.,  et~al., 2020, \mn@doi [\mnras] {10.1093/mnras/staa2648}, \href
  {https://ui.adsabs.harvard.edu/abs/2020MNRAS.499.2896T} {499, 2896}

\bibitem[\protect\citeauthoryear{{Tinker}}{{Tinker}}{2021}]{Tinker2020}
{Tinker} J.~L.,  2021, \mn@doi [\apj] {10.3847/1538-4357/ac2aaa}, \href
  {https://ui.adsabs.harvard.edu/abs/2021ApJ...923..154T} {923, 154}

\bibitem[\protect\citeauthoryear{{Tinker}, {Weinberg}  \& {Warren}}{{Tinker}
  et~al.}{2006}]{tinker2006}
{Tinker} J.~L.,  {Weinberg} D.~H.,   {Warren} M.~S.,  2006, \mn@doi [\apj]
  {10.1086/504795}, \href
  {https://ui.adsabs.harvard.edu/abs/2006ApJ...647..737T} {647, 737}

\bibitem[\protect\citeauthoryear{{Tinker}, {Leauthaud}, {Bundy}, {George},
  {Behroozi}, {Massey}, {Rhodes}  \& {Wechsler}}{{Tinker}
  et~al.}{2013}]{Tinker2013}
{Tinker} J.~L.,  {Leauthaud} A.,  {Bundy} K.,  {George} M.~R.,  {Behroozi} P.,
  {Massey} R.,  {Rhodes} J.,   {Wechsler} R.~H.,  2013, \mn@doi [\apj]
  {10.1088/0004-637X/778/2/93}, \href
  {https://ui.adsabs.harvard.edu/abs/2013ApJ...778...93T} {778, 93}

\bibitem[\protect\citeauthoryear{{Tinker}, {Wetzel}, {Conroy}  \&
  {Mao}}{{Tinker} et~al.}{2017}]{tinker2017b}
{Tinker} J.~L.,  {Wetzel} A.~R.,  {Conroy} C.,   {Mao} Y.-Y.,  2017, \mn@doi
  [\mnras] {10.1093/mnras/stx2066}, \href
  {https://ui.adsabs.harvard.edu/abs/2017MNRAS.472.2504T} {472, 2504}

\bibitem[\protect\citeauthoryear{{Tinker}, {Hahn}, {Mao}, {Wetzel}  \&
  {Conroy}}{{Tinker} et~al.}{2018a}]{tinker2018b}
{Tinker} J.~L.,  {Hahn} C.,  {Mao} Y.-Y.,  {Wetzel} A.~R.,   {Conroy} C.,
  2018a, \mn@doi [\mnras] {10.1093/mnras/sty666}, \href
  {https://ui.adsabs.harvard.edu/abs/2018MNRAS.477..935T} {477, 935}

\bibitem[\protect\citeauthoryear{{Tinker}, {Hahn}, {Mao}  \& {Wetzel}}{{Tinker}
  et~al.}{2018b}]{tinker2018}
{Tinker} J.~L.,  {Hahn} C.,  {Mao} Y.-Y.,   {Wetzel} A.~R.,  2018b, \mn@doi
  [\mnras] {10.1093/mnras/sty1263}, \href
  {https://ui.adsabs.harvard.edu/abs/2018MNRAS.478.4487T} {478, 4487}

\bibitem[\protect\citeauthoryear{{Vale} \& {Ostriker}}{{Vale} \&
  {Ostriker}}{2004}]{Vale2004}
{Vale} A.,  {Ostriker} J.~P.,  2004, \mn@doi [\mnras]
  {10.1111/j.1365-2966.2004.08059.x}, \href
  {https://ui.adsabs.harvard.edu/abs/2004MNRAS.353..189V} {353, 189}

\bibitem[\protect\citeauthoryear{{Watson} et~al.,}{{Watson}
  et~al.}{2015}]{watson2015}
{Watson} D.~F.,  et~al., 2015, \mn@doi [\mnras] {10.1093/mnras/stu2065}, \href
  {https://ui.adsabs.harvard.edu/abs/2015MNRAS.446..651W} {446, 651}

\bibitem[\protect\citeauthoryear{{Wechsler} \& {Tinker}}{{Wechsler} \&
  {Tinker}}{2018}]{Wechsler2018}
{Wechsler} R.~H.,  {Tinker} J.~L.,  2018, \mn@doi [\araa]
  {10.1146/annurev-astro-081817-051756}, \href
  {https://ui.adsabs.harvard.edu/abs/2018ARA&A..56..435W} {56, 435}

\bibitem[\protect\citeauthoryear{{White} \& {Frenk}}{{White} \&
  {Frenk}}{1991}]{White1991}
{White} S. D.~M.,  {Frenk} C.~S.,  1991, \mn@doi [\apj] {10.1086/170483}, \href
  {https://ui.adsabs.harvard.edu/abs/1991ApJ...379...52W} {379, 52}

\bibitem[\protect\citeauthoryear{{Yang}, {Mo}  \& {van den Bosch}}{{Yang}
  et~al.}{2009}]{Yang2009}
{Yang} X.,  {Mo} H.~J.,   {van den Bosch} F.~C.,  2009, \mn@doi [\apj]
  {10.1088/0004-637X/693/1/830}, \href
  {https://ui.adsabs.harvard.edu/abs/2009ApJ...693..830Y} {693, 830}

\bibitem[\protect\citeauthoryear{{Yu} et~al.,}{{Yu} et~al.}{2022}]{Yu2022}
{Yu} J.,  et~al., 2022, \mn@doi [\mnras] {10.1093/mnras/stac2176}, \href
  {https://ui.adsabs.harvard.edu/abs/2022MNRAS.516...57Y} {516, 57}

\bibitem[\protect\citeauthoryear{{Zheng}}{{Zheng}}{2004}]{zheng2004}
{Zheng} Z.,  2004, \mn@doi [\apj] {10.1086/421542}, \href
  {http://adsabs.harvard.edu/abs/2004ApJ...610...61Z} {610, 61}

\bibitem[\protect\citeauthoryear{{Zu} \& {Mandelbaum}}{{Zu} \&
  {Mandelbaum}}{2016}]{Zu2016}
{Zu} Y.,  {Mandelbaum} R.,  2016, \mn@doi [\mnras] {10.1093/mnras/stw221},
  \href {https://ui.adsabs.harvard.edu/abs/2016MNRAS.457.4360Z} {457, 4360}

\bibitem[\protect\citeauthoryear{{Zu} \& {Mandelbaum}}{{Zu} \&
  {Mandelbaum}}{2018}]{Zu2018}
{Zu} Y.,  {Mandelbaum} R.,  2018, \mn@doi [\mnras] {10.1093/mnras/sty279},
  \href {https://ui.adsabs.harvard.edu/abs/2018MNRAS.476.1637Z} {476, 1637}

\bibitem[\protect\citeauthoryear{{de Jong} et~al.,}{{de Jong}
  et~al.}{2014}]{dejong2014}
{de Jong} R.~S.,  et~al., 2014, in {Ramsay} S.~K.,  {McLean} I.~S.,   {Takami}
  H.,  eds,  Society of Photo-Optical Instrumentation Engineers (SPIE)
  Conference Series Vol. 9147, Ground-based and Airborne Instrumentation for
  Astronomy V. p. 91470M, \mn@doi{10.1117/12.2055826}

\makeatother
\end{thebibliography}


\bsp	
\label{lastpage}
\end{document}